\documentstyle[11pt,paspconf,epsf,psfig,twoside]{article} 
 
\def\about{$\sim$}   
   
\def\arcsec{$\,^{\prime\prime}$~}   
\def\la{\hbox{\rlap{$<$}\lower.5ex\hbox{$\sim$}\ }}   
\def\ga{\hbox{\rlap{$>$}\lower.5ex\hbox{$\sim$}\ }}   
\def\Msun{$M_\odot$}   
\def\ref#1{\noindent\hangindent=24.0pt\hangafter=1{#1}\par}   
   
\def\ang{${\rm \AA}$} 
\def\Hbeta{H$\beta$}   
\def\Halpha{H$\alpha$}   
\def\mdot{$\dot{{\rm m}}$ }  
\def\he2{\ion{He}{2}}   
\def\ni{\noindent}   
   
\def\F{$\mathcal{F}$} 
\def\X{$\mathcal{X}$} 
\def\R{$\mathcal{R}$}   
\def\Lx{L$_x$~}

\def\Pb{P$_b$}  
\def\Pp{P$_p$}  
\def\Fx_Fopt{F$_x$/F$_{opt}$}  
 
\begin{document} 

\small 
\vspace*{-15mm}
\begin{flushright}
{to appear in {\it Proc. Annapolis Workshop on Magnetic CVs},\\ 
(C. Hellier and K. Mukai, eds.), ASP Conf. Series, 157 (1999), in press.}
\end{flushright}
\normalsize
  
\title{Magnetic CVs in Globular Clusters}

\author{Jonathan E. Grindlay } 
\affil{Harvard-Smithsonian Center for Astrophysics, Cambridge, MA, U.S.A.}

\begin{abstract} 
Cataclysmic variables (CVs) are tracers of dynamical evolution in 
globular clusters. A significant  sample has been found by the relatively shallow   
\Halpha~ surveys conducted with HST thus far in just three globulars:  
NGC 6397 (3), NGC 6752 (2), and $\omega$-Cen (2). 
These are identified with the low luminosity  
x-ray sources discovered in these globulars with 
similarly moderate-depth ROSAT   surveys. Follow-up spectroscopy of 
the initial 3 CV candidates in NGC 6397 with the FOS on HST  
suggests they may be magnetic CVs of the intermediate polar (IP) type,
and follow-up deep imaging with both HST and ROSAT yield a 4th confirmed 
and 5th strong candidate CV near the center of this collapsed core cluster.  
The optical vs. x-ray properties of the 4 confirmed IP candidates in 
NGC 6397 are  compared with disk IPs, and arguments for and against this  
interpretation are presented. The possible strong excess of magnetic CVs  
in globulars will be tested with much deeper HST/AXAF surveys.
\end{abstract} 
 
\keywords{globular clusters, magnetic white dwarfs, blue stragglers, 
intermediate polars,  DQ Her stars, HeII, truncated accretion disks, 
dim x-ray sources, low mass x-ray binaries (LMXBs)} 
 
\section{Introduction}
 
With the advent of high resolution imaging telescopes in space, the cores  
of globular clusters have become available for studies of compact 
binaries (CBs),  defined (here) as short period (\la1d) systems 
containing a compact object (white dwarf (WD),  neutron star (NS) 
or black hole). CBs are observable as cataclysmic variables (CVs), 
low mass x-ray binaries (LMXBs),  
and millisecond pulsars (MSPs) and play a particularly important  
role in globulars: as the most compact (hard) binaries, they are not  
only the survivors of ``binary burning" which dynamically heat the cluster  
cores while destroying wide binaries, but are the  saviors of the cluster 
against  complete core collapse (cf. Hut et al 1992).  
CVs are likely to be dominant, since white dwarfs (WDs) should vastly  
outnumber neutron stars (NSs) in clusters: for a Salpeter IMF  
with differential mass index \about2.4,    
the WD progenitors (\la8 \Msun) must exceed those for NSs by    
a factor \about(8/0.8)$^{1.4}$ \about25. Conversely, the measured  
fraction, \F,  of CBs containing WDs (CVs) vs.  
NSs (LMXBs + MSPs)  then limits the    
primordial IMF of the cluster and subsequent dynamical evolution    
of its stellar remnants.  A reduction in \F~ is expected, for    
example, by considerations of stable mass transfer and mass   
segregation (e.g. Bailyn, Garcia and Grindlay 1990). Thus 
the study of CVs in globulars constrains both stellar and dynamical 
evolution. 

Not only the number and spatial distribution, but the very nature 
of CVs in globulars is emerging as a clue to their origin. On the 
basis of  HST/FOS spectra (Grindlay et al 1995; GC95) 
showing moderately strong HeII ($\lambda$4686) emission for 
the three \Halpha~ emission objects discovered (Cool et al 1995; CG95) 
in the core of the nearest core-collapsed globular, NGC 6397, 
we have suggested they may be magnetic CVs (MCVs) of the  
DQ Her (hereafter intermediate polar, IP)  type in which the accreting 
WD has a magnetic field B$_{WD}$ \ga 0.1- 1 MG sufficient to truncate 
the inner edge of the accretion disk. More detailed analysis of NGC 6397, 
including a fourth CV candidate found in deep HST imaging 
(Cool et al 1998; CG98) and spectroscopy (Edmonds et al 1999; 
EG99), provides additional evidence that these first 4 spectroscopically 
confirmed CVs in a globular cluster core may be IPs (EG99), 
whereas only \about10\% of CVs in the field are 
of the DQ Her type (cf. Patterson 1994). If confirmed with both more detailed 
optical (and x-ray studies) and larger samples, this would suggest MCVs  
and magnetic WDs are somehow enhanced in globular 
cluster cores, providing yet more evidence that  stellar 
evolution in globulars is affected by close encounters. 
One possibility (Grindlay 1996) is that since rotation of stellar 
cores increases in encounters or mergers, dynamo production 
of magnetic fields may result and the resulting WDs may be 
preferentially magnetic. 

In this paper, we briefly discuss the various search  
techniques for finding (and then studying)  CVs and MCVs in  
globulars. We  summarize the \Halpha-imaging technique  
conducted with HST which has yielded 3 CV candidates 
in NGC 6397 (CG95) and 2 each in NGC 6752 (Bailyn et al 1996; BR96)  
and $\omega$-Cen (Carson et al 1999). We focus on NGC 6397 for 
which the initial spectroscopic followup studies  
(GC95) as well as UBVI photometry revealed a fourth CV (CG98). 
We compare the optical emission line and continuum spectra  
of the 4 CVs in NGC 6397 with correlations found for field  
CVs and find they are consistent with those for IPs.  
Using the new spectrophotometry of EG99, and the preliminary 
results of  our follow-up deep (75ksec) ROSAT observation 
(Grindlay, Metchev and Cool 1999; GMC99), 
we compare the x-ray vs. optical properties of  
the brightest 4 spectroscopically confirmed CVs and find  
they are consistent with  the optical vs. x-ray correlations displayed 
by IPs in  the disk. We report a fifth CVcandidate in 
the cluster core from the 
deep ROSAT study and find a likely optical counterpart 
as a near uv-excess star measured by CG98 that would also 
be consistent in its x-ray/optical flux ratio. Although the dim x-ray sources 
in NGC 6397 resemble IPs, we also briefly reconsider  whether they might 
instead be very low accretion  rate and optically thin disks as expected  
for either quiescent dwarf novae containing WDs or quiescent LMXBs 
containing NSs. 
 
We conclude with a brief comparison of our current results for NGC 6397 with  
those for 47 Tuc, for which Verbunt and Hasinger (1998; VH98)  
have a moderate ROSAT exposure. Interesting differences  and 
similarities in the CV and CB populations may already be apparent.  
Upcoming HST and AXAF observations will provide much more sensitive 
tests of  the possible MCV excess in globulars.   
 
\section{CV Searches in Globulars} 

CVs have long been sought in globular clusters  
for a variety of (good)  
reasons: known distances to then fix mass transfer rates \mdot; Pop II  
environments to test halo models; formation histories likely to include  
stellar encounters to contrast with primordial binary evolution (only)  
for field objects; and many others. Here we briefly review the search  
techniques and their completeness for finding MCVs.  
 
\subsection{UV-excess and Variability} 
Since most CVs have been discovered in the field as blue  
variables, with  dwarf novae (DNe) being the most common and  
novae the most extreme examples,  initial searches in  
globulars have emphasized these properties. Indeed the  
only pre-HST spectroscopically confirmed CV in a globular (V101 in M5;  
Margon et al 1981) is a DN as originally suggested by  
Oosterhoff  on the basis of outbursts. We note below that  
the total CV (and perhaps CB) population in globulars is strongly  
centrally concentrated so that V101, which could never be detected  
(from the ground) in the core, requires either  formation from a  
primordial cluster binary or ejection from the core. 
 
The cores of two clusters have been moderately well  
searched with HST imaging for blue variables: NGC 6752  
(Shara et al 1996) has yielded only upper limits for DNe,  
whereas 47 Tuc has produced only one confirmed and one  
possible DN system (Paresce and de Marchi 1994). Although these 
searches  have certainly been sensitive to DN outbursts, the ability to  
detect \about0.1-0.2 mag flickering typical of quiescent  
DNe (with apparent magnitudes \ga21 for  
either cluster) is questionable for blind searches (though less so for  
identified objects, where neighbor subtraction can be  
accomplished more reliably). The fact that the CV candidates  
in NGC 6397 and NGC 6752 have now been found to be so  
red that in V-I they are nearly on the cluster main sequence (CG98),  
though still with  moderate U-B excess,   
also suggests that uv-excess alone is not a requirement for cluster  
CVs. This is reinforced by the relatively uv-deficient spectral  
distribution of the one cluster CV studied now in the far-uv, CV1 in  
NGC 6397 (EG99).  If the cluster CVs are dominated by MCVs 
then both uv-excess and dwarf nova  
type variability are expected to be suppressed due to the  
truncation of the inner portion of the accretion disk. Thus  
CV searches should be constructed to be independent of  
these criteria. 
 
\subsection{\Halpha~ Emission Imaging} 
All CVs except some DNe in outburst show emission lines,  
with the Balmer lines and \Halpha~ most intense. This  
motivated our search strategy, developed initially in  
relatively shallow ground-based searches (e.g.  
Cool 1993) and culminating with HST and the initial  
results for NGC 6397 (CG95). An emission line search using  
narrow-band imaging would ideally use three filters: one centered  
on the emission line and two flanking continuum filters to 
measure both local continuum and slope  
while averaging over adjacent absorption line features 
(as used in our CTIO search of NGC 6752  (cf. Cool 1993)). Given the  
WFPC filter set, we have chosen the available \Halpha~  
filter F656N (with width W$_L$ \about 20 \ang) and a single broad  
continuum filter (F675W with width W$_L$ \about 913 \ang;   
yielding \R~ magnitudes). \Halpha~ emission  
candidates are then identified in the color magnitude diagram formed  
by \R~ vs. \Halpha~ - \R~ as ``blue'' objects. Calibration of  
this photometry is self contained by the measure of  blue stragglers  
and horizontal branch stars, with their stronger \Halpha~ absorption,  
which show up as a separate track of ``red'' objects offset by  
\about0.15mag in the CMD (cf. CG95).   
 
If MCVs dominate, the \Halpha~ searches should be relatively sensitive.  
However, the searches are per force limited by the narrow-band  
throughput of the F656N filter and consequent long total effective  
integrations to achieve interestingly deep limiting magnitudes.  
In NGC 6397, the closest globular with a high density core, we  
shall obtain (in cycle 7) 15 HST orbits to reach an effective limiting  
magnitude of \about24 (M$_V$ \about 11.5;\about10$\sigma$).
This  would span \ga3-4 binary  orbits of the \la6h expected orbital periods 
(given the limits on  secondary mass as well as disk absolute magnitudes  
derived by EG99) of the cluster CVs.  However, once identified in \Halpha,  
variability studies can be conducted and the likely  
orbital periods \Pb~  (e.g.  \about3.5 - 5h for  CVs 1-4 in NGC 6397; 
cf. EG99)  can be measured from the accompanying short R exposures 
as done successfully for the 2 CV candidates in NGC 6752 (BR96).  
Detection of pulsation periods \Pp \la \Pb~ would provide  direct 
confirmation that they are indeed IPs and may be  possible for the brightest  
candidates by searching for modulations in the B continuum  
by temporal analysis of STIS spectra as we have proposed.  
 
\subsection{Dim X-ray Sources} 
 
A new population of dim x-ray sources was discovered in 
globular cluster cores and   
proposed as most likely to be CVs (Hertz and Grindlay 1983; HG83). 
Given the usual  advantage of known distances to globular clusters, 
the luminosities of  these dim sources could be derived with greater accuracy 
than for  field CVs (although the fainter source fluxes precluded spectral  
determinations). Since the initial (relatively shallow)  Einstein surveys yielded  
luminosities \Lx(0.2-4keV) \about 10$^{32.5 - 34.5}$ erg/s, and since  
at least one of the dim sources (in NGC 6440) was regarded as the likely  
detection of a quiescent NS transient, HG83 concluded the dim sources  
were likely a mixture of both CVs and quiescent  LMXBs (qLMXBs).  
Verbunt et al (1984) argued that all the dim sources discovered  
with the Einstein survey were most likely  
qLMXBs although it was already evident from studies of field CVs  
(e.g. Patterson and Raymond, 1985a,b; PRa,b) that the luminosities of  
field CVs extended above \about10$^{32.5}$ erg/s in the 
0.5-4.5 keV Einstein band.  
 
Much greater sensitivity surveys have now been conducted with ROSAT.  
Here we only consider the initial  
(shallow) survey of NGC 6397 (Cool et al 1993; CG93)  
and the current (deepest) results on NGC 6397 (GMC99)  
and 47 Tuc (VH98). CG93 discovered  three  
sources with \Lx \about 10$^{31.5-32}$ erg/s within \about 10\arcsec of the  
center of NGC 6397. These comfortably overlap typical CV luminosities, and 
indeed  our initial \Halpha~ imaging survey of NGC 6397 with HST (CG95) 
revealed  three optical candidates, with spectra showing them to be most  
likely IPs (GC95). The deeper survey of GMC99 shows  at least a 4th source 
(and probably several more)  in the central core, as discussed  
below. MCVs might be  expected to dominate the ROSAT survey  
since they typically have \Fx_Fopt values greater  
than non-magnetic CVs (PRa, Patterson 1994, Beuermann 1998).    
However, the \Halpha~ survey found (blindly) the same sources  within the 
expected sensitivity limits suggesting that any possible MCV excess in  
globulars is not a result of x-ray selection alone.

\section{Evidence for MCVs} 
 
MCVs in globulars were suggested as a possible observable class  
by Chanmugam, Ray and Singh (1991). The first evidence for their  
detection was contained within the HST/FOS spectra of CV  
candidates 1-3 in NGC 6397 which showed all to have moderately strong  
HeII emission (GC95).  Although HeII ($\lambda$4686) emission is not unique 
to MCVs, and  in fact PRb show that it is correlated with accretion rate \mdot  
and present with EW(HeII) \about 3\ang~ in most CVs, Silber (1992)  
has shown that the apparent excitation as measured by the  
ratio of equivalent widths \X~ = EW(HeII)/EW(\Hbeta)   
correlates with magnetic nature, with \X~ \ga 0.3 for intermediate  
polar (IP; or DQ Her  type) or polar (or AM Her type) systems.  
Although CVs  1 - 3 in fact have \X~ = 0.32, 0.34, and 0.25,  
and CV 4 has \X~ = 0.07, EG99 show that their spectra (cf. Figure 1) 
and continuum  properties are in fact most consistent with IPs.  
 
Why should the MCVs have enhanced HeII  emission ? The likely  
reason is the larger optically thin coronal region inside  
the inner edge of a magnetically truncated accretion disk, which  
is then more readily photoionized by soft x-ray emission from  
the accretion column onto the WD. Doppler imaging maps of  
HeII vs. \Hbeta~ support this general picture.  


\psfig{file=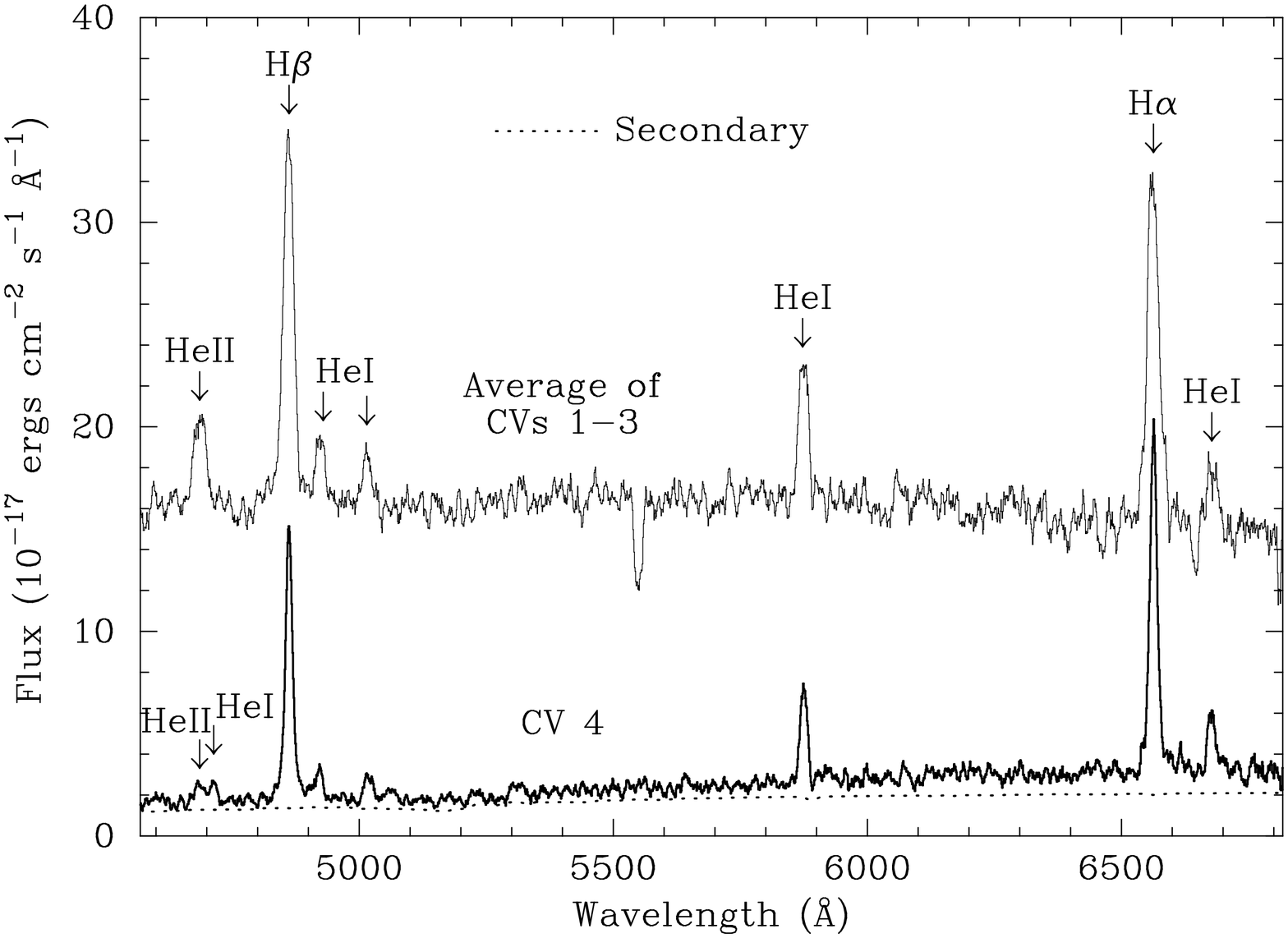,width=11cm,height= 8cm,angle=-90.} 

\noindent
Fig. 1. {Mean spectra of CVs 1-3 (GC95) and CV4 in NGC 6397 (from  
EG99) showing the higher excitation HeI and  HeII emission  
(particularly CVs 1-3) expected for IP type MCVs. (the 
feature at \about$\lambda$5560 is instrumental).} 


\subsection{Colors and Spectrophotometry with HST} 
 
EG99 have investigated the question of how the four CV  
candidates in NGC 6397 (the only spectra available for  
CVs in the cores of globulars)  
compare in both their continuum and line ratio properties  
with CVs generally. Since the V-I colors of these objects  
are nearly coincident with the main sequence for the cluster  
(CG98), reasonably accurate magnitudes and masses for  
the secondaries can be derived (CG98, EG99) and the disk  
absolute magnitudes determined. From plotting these disk  
magnitudes and the continuum ratios at \Hbeta/\Halpha~  
vs. the excitation parameter, \X, the objects closely resemble  
correlations in these quantities obeyed by MCVs, as  
seen in Figure 2. 
 

\begin{minipage}{65mm} 
\psfig{file=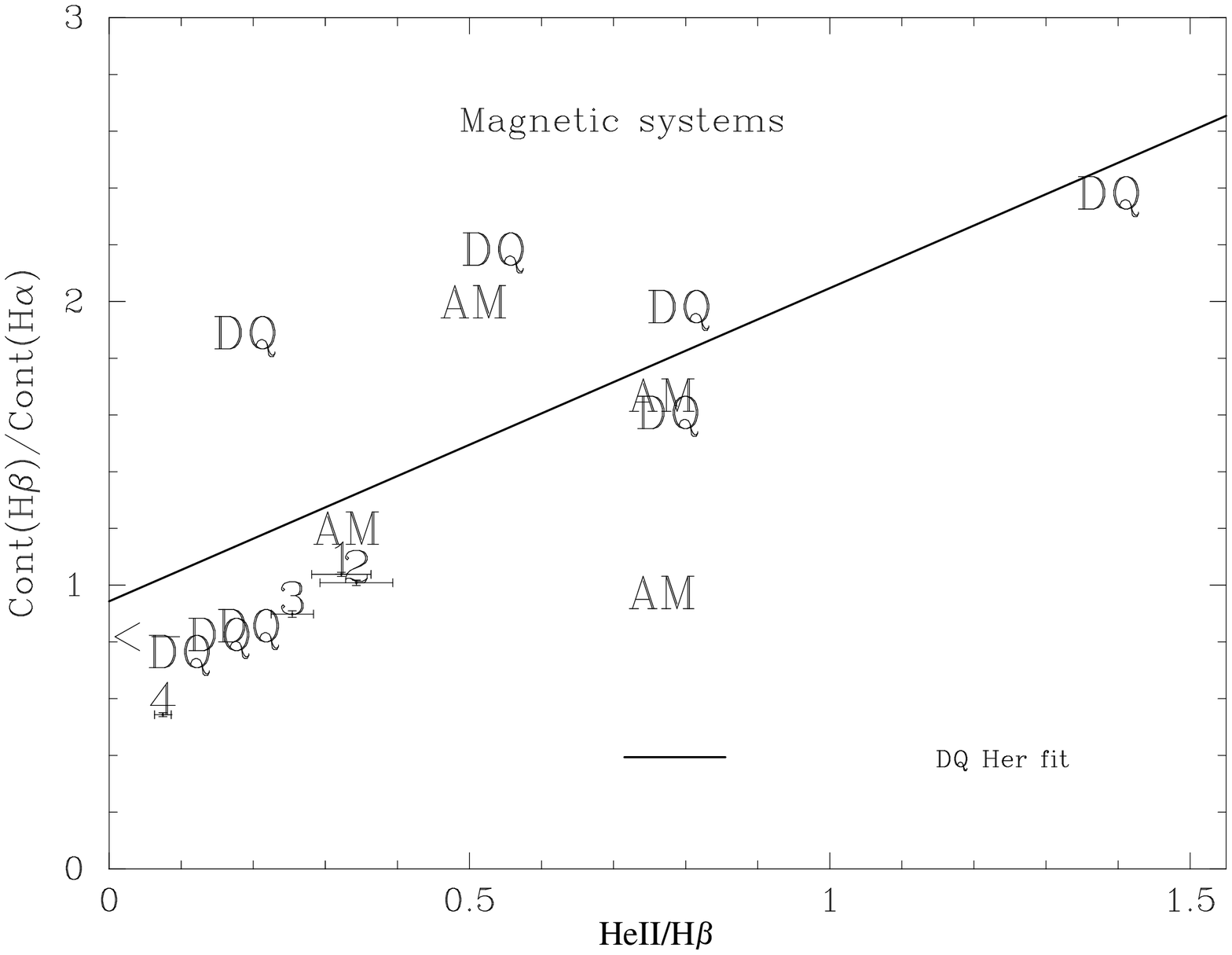,width=55mm} 
\end{minipage} 
\hspace*{1cm} 
\begin{minipage}{65mm} 
\psfig{file=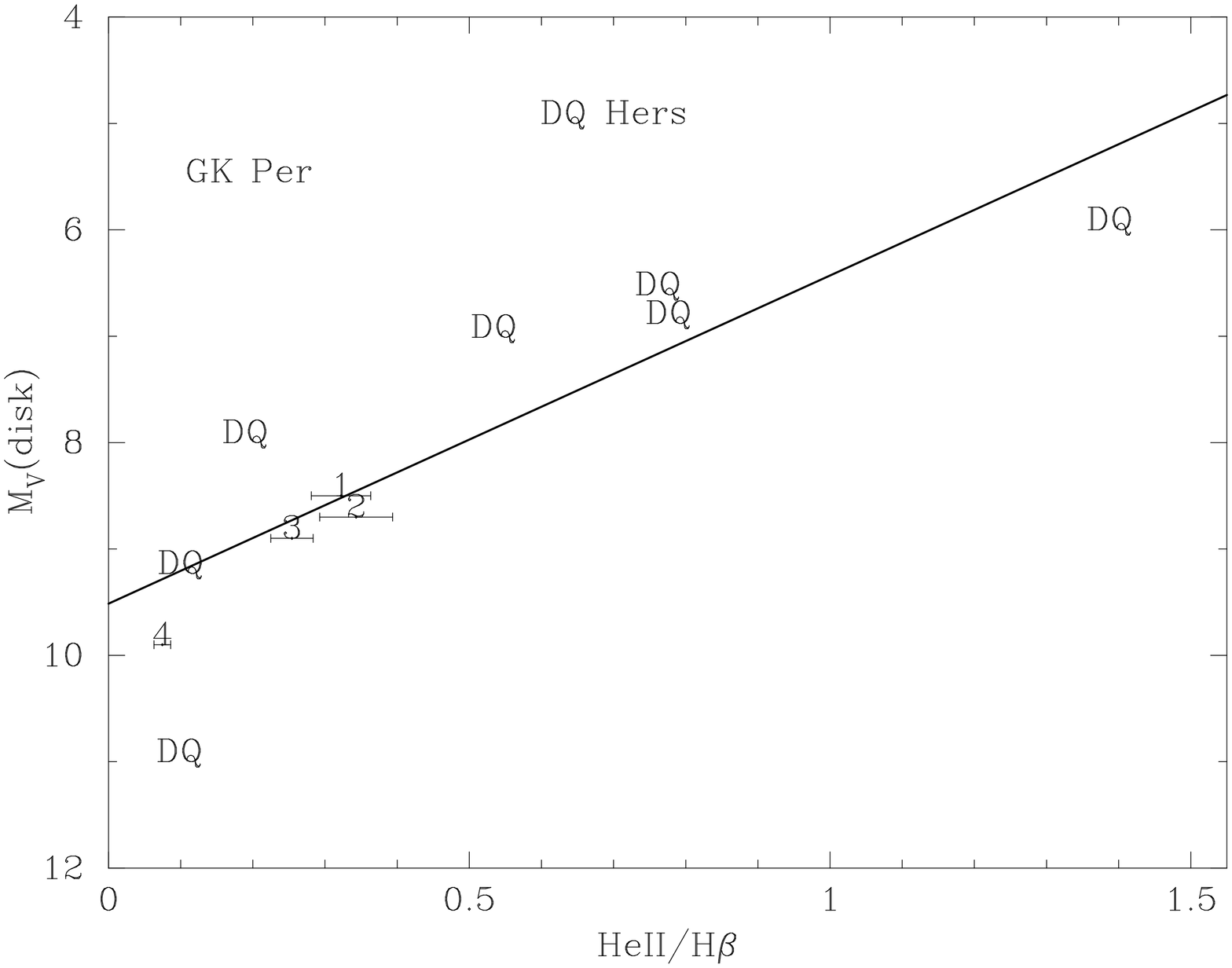,width=55mm,angle=-90} 
\end{minipage} 

\vspace*{5mm}

\noindent
Fig. 2. {Correlations of continuum fluxes at \Hbeta~ vs. \Halpha~ 
and derived disk absolute magnitude vs. the excitation ratio, \X,  
which support the hypothesis that CVs 1-4 in NGC 6397 are  
DQ Her types (from EG99).} \\

 
In Figure 3  we explore the apparent relationship between \X~  
and  EW(\Hbeta), which itself is proportional  
to the relative x-ray to optical flux (see below), for the 4 CVs in NGC 6397 
as well as  7 IPs in the field.  
The emission line data (EW values for HeII and \Hbeta) for the  disk IPs are 
taken  from Williams (1983) (for GK Per, EX Hya, YY Dra, FO Aqr and 
AO Psc),   Steiner et al (1981) (V1223 Sgr) and Motch et al (1996) (V709 Cas)  
and from EG99 for the four CVs in NGC 6397. This emission line  
``CMD'' shows that \X~ is (weakly) anti-correlated with EW(\Hbeta).  
This could reflect a variation in inner disk radius (from either  
WD magnetic field, B$_{WD}$, or accretion rate, \mdot)    
if the HeII emission from within the inner disk  
increases less with increasing accretion rate \mdot than does \Hbeta~   
from the outer disk. 
Although not noted, a  similar correlation may be  
inferred from the data plotted by Echevarria (1988).  
The cluster CVs could define the lower B field 
end of the sequence, since the alternative of higher \mdot 
(alone) is not consistent with their relatively faint disk 
absolute magnitudes (cf. Figure 2). 
Comparison with \Pb~ and \Pp~ values given 
by Hellier (1996) reveals no correlation. 

 
\vspace{-1.cm} 
\hspace{1cm} 
\psfig{file=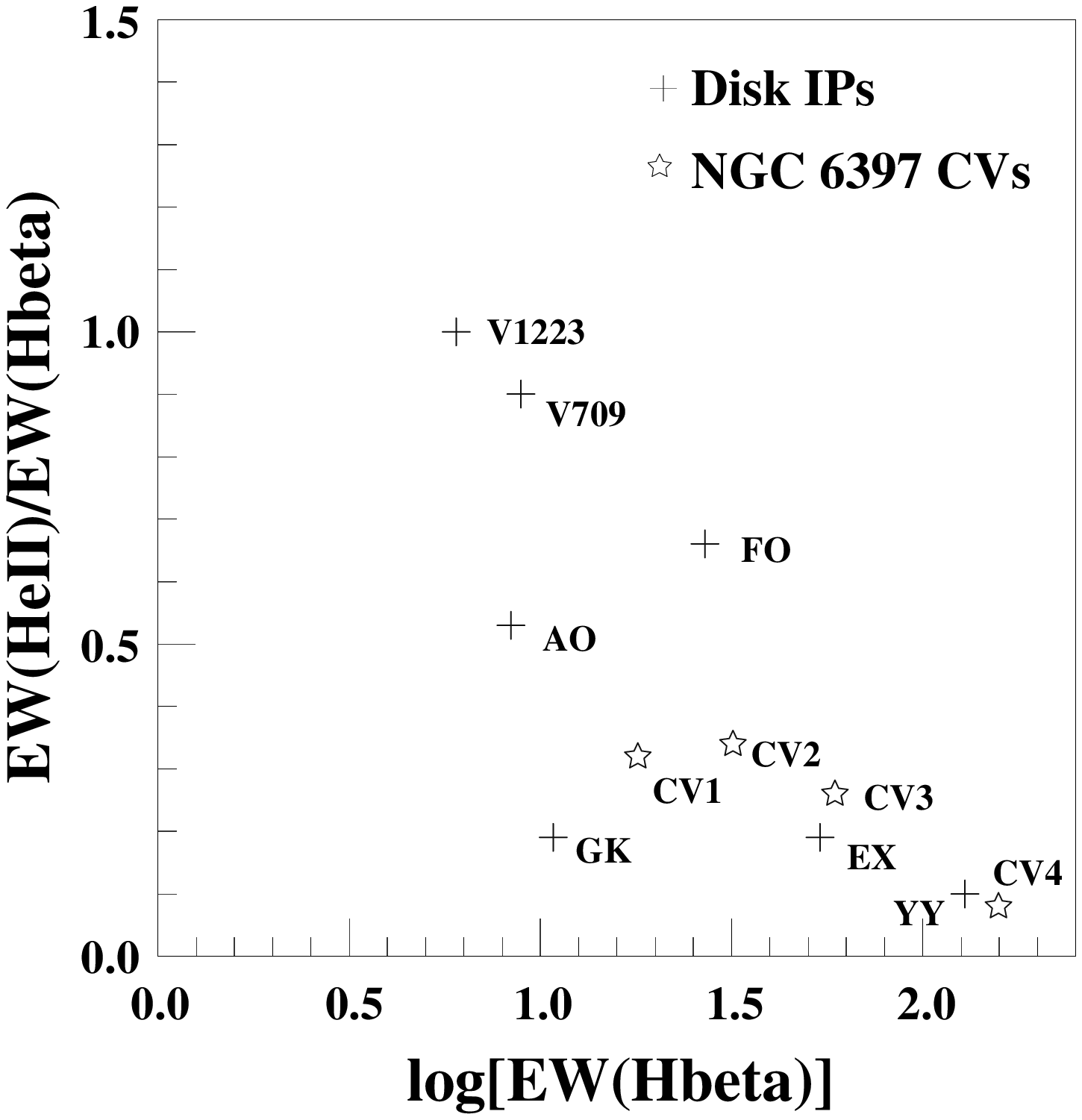,width=10cm,height=9cm} 
 
\vspace{-2.5cm} 
 
\noindent 
Fig. 3: {Excitation ratio, \X, vs. EW(\Hbeta) for  
CVs 1-4 in NGC 6397 vs. disk IPs}. 
 

\subsection{X-ray vs. Optical Properties of CVs in NGC 6397} 
 
Additional tests for the MCV nature of the globular  cluster CVs are possible 
by  comparing their x-ray vs.  optical properties with those of IPs in the disk.  
Using the initial ROSAT x-ray fluxes of the three dim sources, C1-C3, 
in the central core of NGC 6397 (CG93)  and their probable optical 
counterpart CVs 1-3 (cf. CG95  for associations), CG95 found these 
three objects were consistent with the distance-independent 
correlation between \Fx_Fopt and EW(\Hbeta) found by PRa for disk CVs. 
Using the actual measured EW(\Hbeta) 
values (rather than inferred from \Halpha~ magnitudes, as in CG95), 
Grindlay and Cool (1996) refined this correlation and compared 
CVs 1-3 with the qLMXB Cen X-4 (cf. discussion below). Here we 
use our more accurate spectrophotometry given in EG99 and 
preliminary results from our deep (75ksec) ROSAT/HRI 
observation (GMC99). 

We find at least one additional dim source, C4, in 
the core of NGC 6397 as well as additional fainter sources near the core. 
C4 is \about8\arcsec due west of C3 (=CV2) and in the 75ksec 
ROSAT/HRI observation had total flux \about80 counts vs. \about160, 
70 and 150 counts for dim sources C1-C3, respectively. 
Since the positions for CV1 and CV4  
(CG98) are only \about3\arcsec apart, they are not resolved  
by the HRI (with \about5\arcsec resolution) and we divide  
the detected counts for the  dim source C2 between them. The  
derived positions for C1-C3 are each within \about2\arcsec  
of the positions for CVs 1-4, lending confidence to the 
optical identifications and allowing us to search for the 
possible counterpart of the new source, C4. Re-examination 
of the HST/UBVI images reported in CG98 in fact yields  
a very probable identification for a 5th(!) CV in 
the core of NGC 6397: candidate CV5, with apparent magnitude 
V = 21.7 and (U-B) = -0.8 (visible in Figure 3 of CG98) is 
within the \about3\arcsec error circle of dim source C4. 
This star was too faint to be detectable in our original 
\Halpha~ search (CG95) but with \Fx_Fopt \about 3.2, it should be 
easily detected in our forthcoming deep (HST cycle 7) \Halpha~ survey 
of NGC 6397, given the strong correlation between \Fx_Fopt and 
EW(\Hbeta) (PRa). 
 
In Figure 4 we show the derived relation between  
the ratio of x-ray (ROSAT band) to optical (V band) fluxes,  
\Fx_Fopt, vs. EW(\Hbeta). Only CVs1-4 (with 
measured EW(\Hbeta) values) are shown, along with the same  
field IPs as plotted in Figure 3, so that their optical vs.  
x-ray properties may be compared. The flux ratio \Fx_Fopt 
has been computed for each object as the measured flux 
in the V band (5000 - 6000\ang) and the ROSAT band 
(0.5 - 2.5 keV) in order to use measured (not extrapolated) 
values. We use the visual magnitude 
without interstellar reddening so that the NGC 6397 CVs, 
with measured cluster extinction of A$_V$ = 0.58 (cf. CG98) 
may be more properly compared with the disk CVs which 
are all within (typically) 500 pc and only moderately 
reddened. If this correction is not made for the cluster CVs, 
(e.g. if some disk CVs are also reddened), their flux ratios 
would increase by \about0.2 on the log scales plotted in 
the figures below. The x-ray fluxes have been computed for 
all objects assuming a relatively hard bremsstrahlung 
spectrum with kT = 10 keV since this is generally appropriate 
for disk IPs (cf. Patterson 1994), and with an absorption 
column of NH = 1. $\times$ 10$^{21}$ cm$^{-2}$. This NH is the 
interstellar value for NGC 6397 and thus a lower limit since 
disk IPs generally appear to be self-absorbed with NH values 
well in excess of interstellar values (Patterson 1994, Hellier 1996). 

A measure of the uncertainty in F$_X$ from both NH and spectral 
differences for the disk IPs can be obtained by 
comparing the ROSAT 0.5-2.5 keV fluxes (from PSPC survey fluxes given 
by Verbunt et al 1997 for all but YY Dra and V709 Cas, for 
which HRI fluxes from Norton et al 1998 are used) with extrapolating 
the 2-10 keV fluxes and spectral fits given by Patterson (1994) 
(for all but V709 Cas) into the 0.5-2.5 keV band. The error 
bars on the \Fx_Fopt plots denote this hard vs. soft flux
spectral uncertainty, and the line plotted is \\
 
log(\Fx_Fopt) = -2.21~ +  ~1.45 log [EW(\Hbeta)]\\ 
 
\noindent 
as found by PR85a for all CVs. Note that only 4/7 of the disk IPs are above 
the line and that the \Fx_Fopt 
values for CV1 and CV4 are particularly uncertain.

 
\vspace{-0.5cm} 
 
\psfig{file=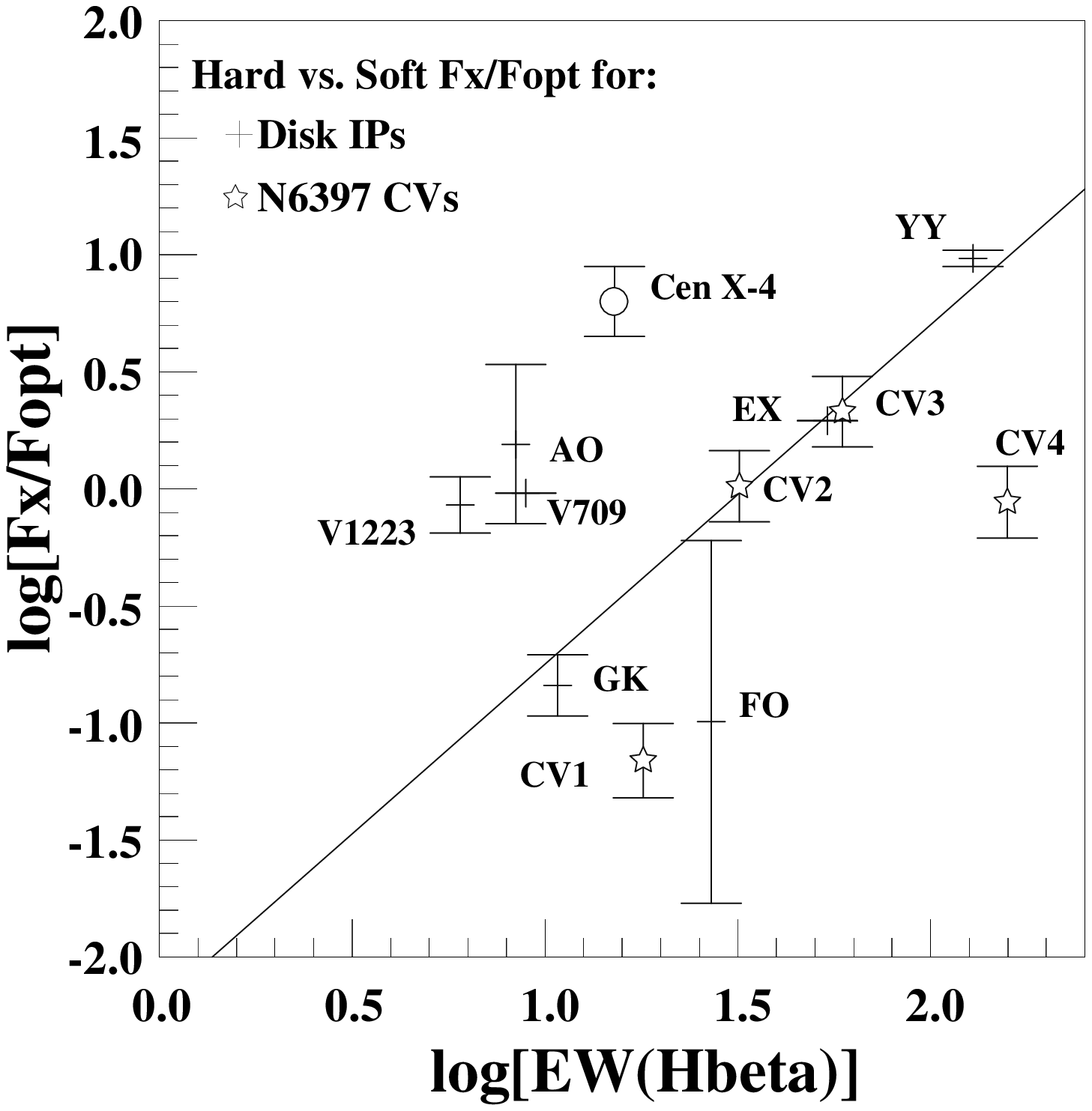,width=10cm,height=9cm} 
 
\vspace{-2.2cm} 
 
\ni 
Fig. 4: {X-ray/optical flux ratio vs. EW(\Hbeta) values for  CVs 1-4 in 
NGC 6397 vs. disk IPs compared with PR relation  for field CVs as  well 
as Cen X-4.}\\ 
 
   
For comparison we show in Figure 5  the same relation   
for HeII vs. \Fx_Fopt. 
 
 
\vspace{-0.5cm} 
 
\psfig{file=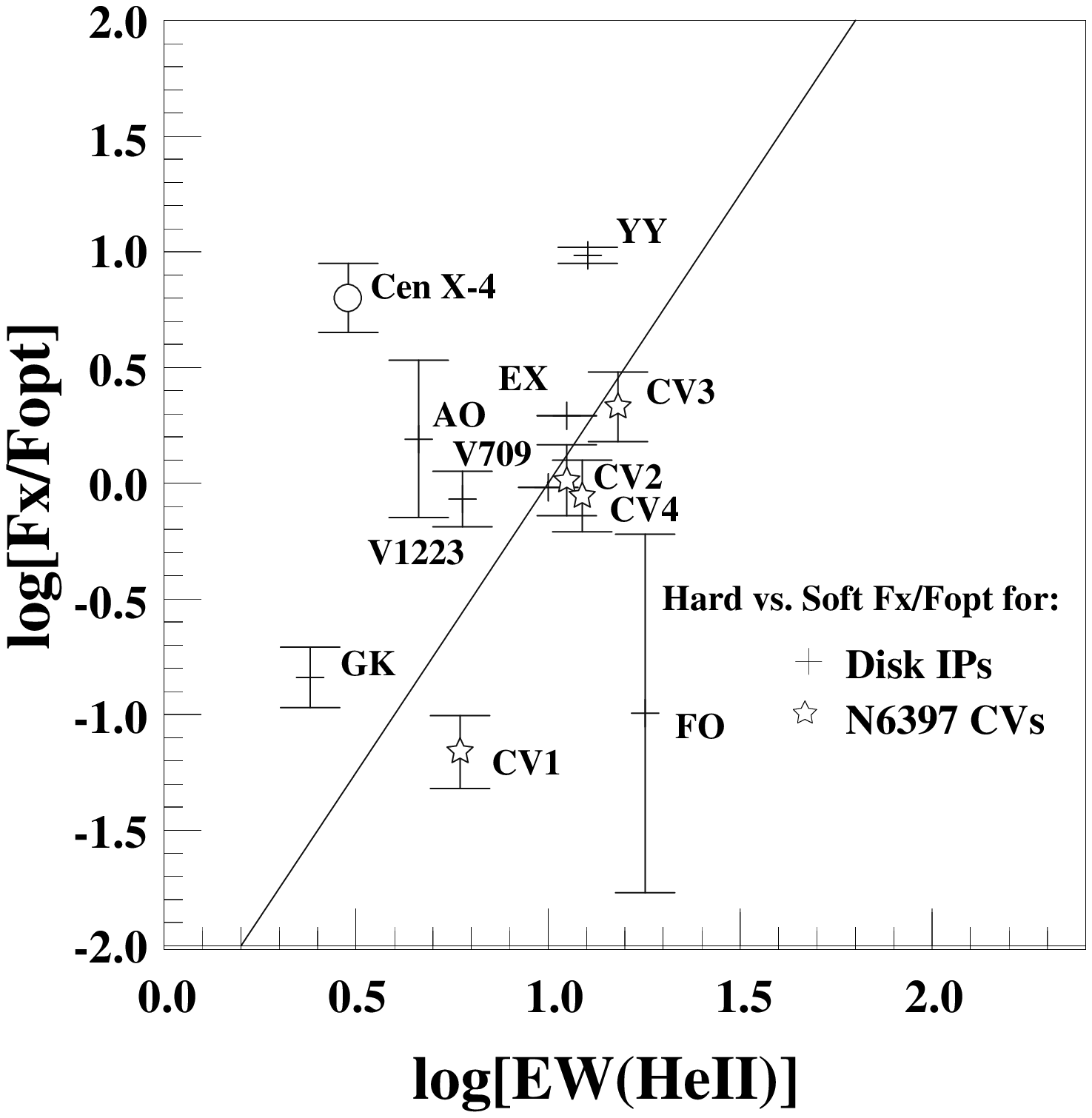,width=10cm,height=9cm} 
 
\vspace{-2.2cm} 
 
\ni 
Fig. 5: {\Fx_Fopt vs. EW(HeII) and approximate fit.}\\

 
\noindent 
The approximate linear fit to the log-log  data plotted  
is given by\\ 
 
log(\Fx_Fopt) = -2.5 ~ + ~ 2.5 log [EW(HeII)]\\ 
 
\noindent 
Thus the x-ray/optical flux is more strongly dependent  
on the HeII line strength than on \Hbeta.  
 
Finally, we investigate the possible relation between \X~  
and \Fx_Fopt directly since the correlations of \Fx_Fopt 
with both EW(\Hbeta) and EW(HeII) might at first suggest 
a positive correlation with \X. However, algebraically, the 
approximate log-log relations given above would predict 
\X \about (\Fx_Fopt)$^{-4/15}$, which is plotted 
in  Figure 6 with the same data points.  

 
\vspace{-0.5cm} 
 
\psfig{file=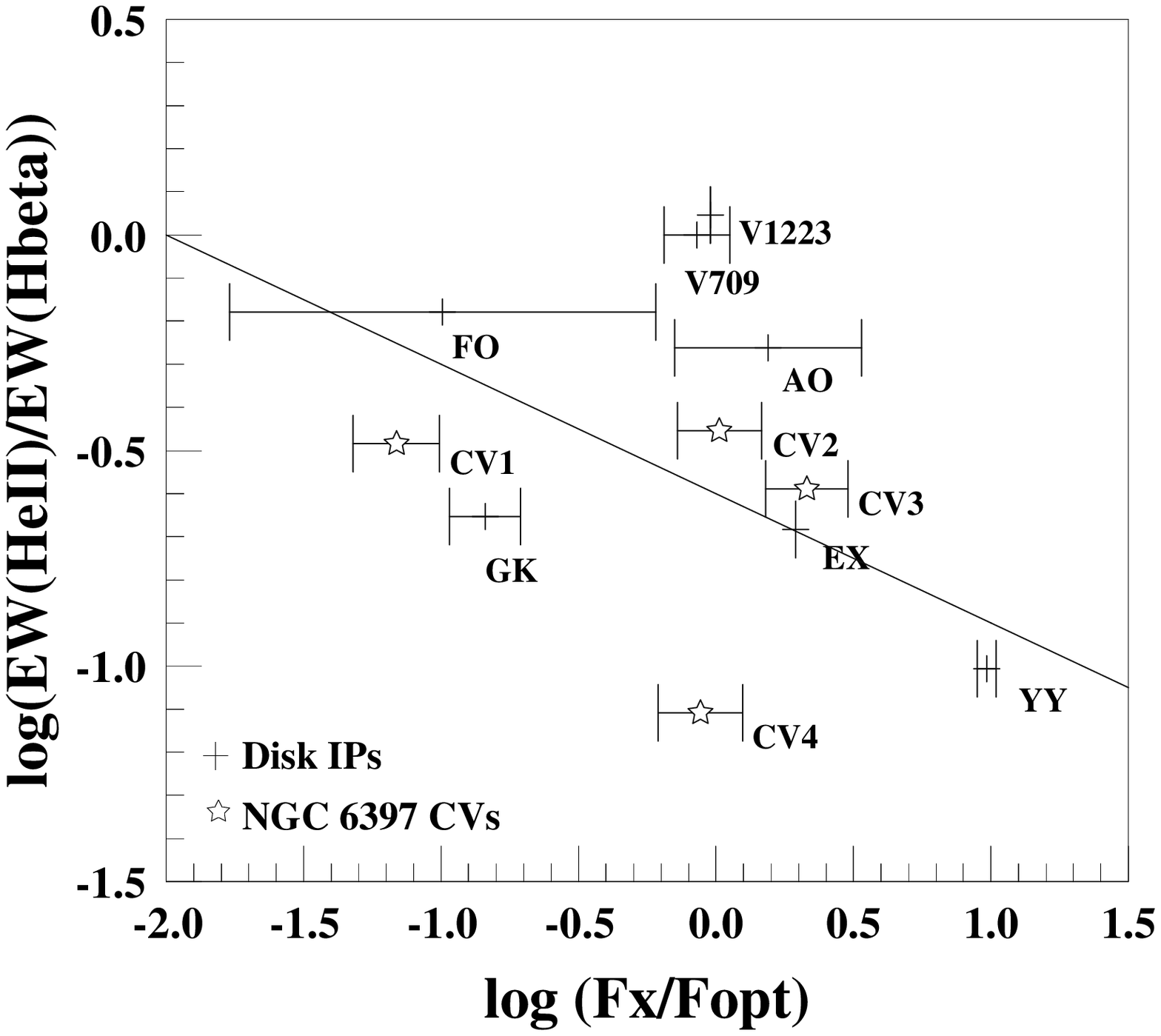,width=10cm,height=9cm} 

\vspace{-2cm} 
 
\ni 
Fig. 6: {Excitation ratio, \X, vs. \Fx_Fopt for  
CVs 1-4 in NGC 6397 vs. disk IPs.}\\  
 
 
\noindent 
We note that since both CV1 and  
GK Per have their \Fx_Fopt ratios most strongly affected  
by their relatively massive secondary companions, their  
\Fx_Fopt ratios in Figures 3-6 may be anomalously low. 

\subsection{Are the Objects in NCG 6397 Really MCVs ?} 
The spectra, photometry and x-ray/optical properties 
of CVs 1-4 in NGC 6397 (cf. EG99) are consistent with them 
being MCVs (IPs), yet until pulsations or other signatures 
unique to IPs are detected there are still questions:\\ 
 
\noindent 
{\it Are they quiescent Dwarf Novae ?} 
Their  faint disks, and the apparent lack of DNe generally  
in globulars (Shara et al 1996) could both be indicative of  
WZ Sge type systems in which the recurrence time for  
DN outbursts has become very long at the low accretion  
rates possibly implied by the faint (optically thin) disks.\\   
 
\noindent
{\it Are they quiescent LMXBs ?}
In Figures 4 and 5 we have plotted the \Fx_Fopt and EW(\Hbeta), 
EW(HeII) values, respectively, for the classic qLMXB Cen X-4. 
The x-ray flux is from the recent ASCA spectral measurement 
(Campagna et al 1997) and the optical magnitudes and EW 
values are from Chevalier et al (1989) and McClintock and 
Remillard (1990). It is clear that Cen X-4 is offset from 
the bulk of the IPs in both correlations, and even moreso 
from CVs1-4 (the forthcoming measurement of EW(\Halpha) \about 
EW(\Hbeta) for CV5, with log(\Fx_Fopt) \about 0.5, will provide 
an additional test). However, since the qLMXB x-ray 
luminosities (e.g. Cen X-4) are \ga10$\times$ larger than 
the \about1-4 $\times$ 10$^{31}$ erg/s values for CVs1-4 (and also 
CV5), a qLMXB interpretation is less likely.

\section{Discussion}
It is striking that CVs 1-5 are all within \about7\arcsec (\about0.08pc) 
of their centroid position in NGC 6397, which is itself 
\about3\arcsec NW of the position given by Sosin (1997) for the cluster center. 
The radial extent of these central CVs is comparable with the 5\arcsec 
core radius derived by Sosin (1997) for this post core collapse (PCC) cluster as 
well as for the distribution of bright central blue stragglers (BSs) noted by 
Auriere et al (1990). This may support the suggestion (Grindlay 1996) 
that the required magnetic WDs in cluster cores (if indeed the CVs 
are IPs) are produced in BSs. If the core is in equipartition and BSs 
have masses \about2$\times$ the turnoff value or \about1.5\Msun, the 
implied WD masses in CVs1-4 are \about1\Msun~ given their 
\about0.5\Msun~ (EG99) secondaries.

The distribution of dim 
sources in 47 Tuc reported by VH98 is similar: the central 5 are within 
\about20\arcsec of the cluster center, or again comparable to the cluster 
core radius and BS distribution of this non-PCC cluster, although the dim sources 
are each typically \about10-50$\times$ 
more luminous (the brightest may be qLMXBs). The underlying extended emission 
in the core quoted by VH98, with total 
luminosity \about4 $\times$ 10$^{32}$ erg/s, 
is about twice the total core luminosity of NGC 6397 and may reflect a similar 
distribution of (\about10) fainter CVs. Since the core of 47 Tuc contains  
\ga10$\times$ the mass of the NGC 6397 core, the nearly comparable numbers of CVs 
suggest the core collapse may have triggered a burst of  CB production in NGC 6397. 
If so, it is remarkable that NGC 6397 as yet contains no compelling evidence 
for CBs containing NSs: no MSPs have yet been reported (despite at least two 
surveys) whereas at least 11 are known in 47 Tuc. This may reflect differences 
in the cluster IMFs or NS retention. 

Upcoming high resolution x-ray imaging 
and spectra with AXAF of both clusters, and the deep \Halpha~ survey of 
NGC 6397 (with 47 Tuc still needed), 
will help measure the CV nature and content. AXAF, in particular, 
will resolve CVs 1 vs. 4 in NGC 6397 (thus removing 
the uncertainties in their \Fx_Fopt
values as plotted in Figures 4 and 5) as well as the fainter sources 
in both clusters. ACIS spectra can also test for whether the sources 
have the hard spectra typical of IPs. However, deep STIS spectra and 
temporal analysis for pulsations are also needed to clarify if the objects 
are indeed dominated by IPs. 

\acknowledgments 

I thank A. Cool, P. Edmonds, and S. Metchev for assistance with analysis and 
HST grant GO-6742 for partial support.

\end{document}